\documentclass{aa}
\usepackage{graphicx}
\begin{document}

\title{Discovery of Two New Faint Cataclysmic Variables\thanks{Based on 
observations obtained at ESO La Silla (ESO Proposal 69.D-0142(A))}}

\author{S. B. Howell \inst{1}, E. Mason \inst{2}, M. Huber \inst{1} 
\and
R. Clowes \inst{3}}

\offprints{S. Howell}

\institute{
Astrophysics Group, Planetary Science Institute, 
Tucson, AZ
\and
ESO, Alonso de Cordova 3017, Casilla 19001, Vitacura, Santiago, Chile
\and
Computational Astrophysics, Department of Computing, 
University of Central Lancashire, Preston PR1 2HE
}

\date{Accepted to A\&A Letters}

\abstract{
We report on the discovery of two new faint cataclysmic variables. 
The objects were selected as candidates from two different 
imaging surveys aimed at the discovery of such faint systems.
One survey used color and variability while the other used color and H$_\alpha$
emission as selection criteria. We present our spectra of the two new variables
and discuss their properties. A discussion of the implication of these 
discoveries on the space density of faint cataclysmic variables is presented. 
}

   \titlerunning{New CVs}
   \authorrunning{S. Howell et al.}
   \maketitle
%
%________________________________________________________________

\section{Introduction}

Cataclysmic variables (CVs) are an important class of binary stars 
as they contain 
one of the most populous stars in the universe, a white dwarf. 
The other star in a CV is a low mass object, generally 0.8-0.1 M$_\odot$,
which loses mass via Roche Lobe overflow toward the more massive primary.
A review of the general properties of CVs is given in Warner (1995).

Recent theoretical work (e.g., Howell et al. 2001, Kolb 2001) 
has predicted that the majority of 
present day CVs will have very short orbital periods ($<$2 hours) and be of low
luminosity (M$_V<$11), i.e., they will be TOADs (Tremendous Outburst Amplitude Dwarf
novae; Howell et al. 1995). These predictions are based on binary evolution models
using the accepted, long held paradigms of CV evolution theory. However, recent new
ideas related to the level at which magnetic fields drain orbital angular 
momentum, 
and if and when they cease to be important for the mass donor star,
have been proposed
(King et al. 2002; Howell et al. 2000). Additionally, the idea of circumbinary
disks as an orbital angular momentum sink has been 
proposed (Taam and Spruit 2001,
2002), observationally searched for (Singha et al. 2002) and possibly
discovered (Mouchet et al. 2002).
If these new theoretical ideas are shown to be correct, the predicted plethora of short 
period, faint CVs
may not really exist. However, it may be that the oldest systems 
no longer resemble CVs at all (e.g., Harrison et al. 2002). 

Other imaging surveys with spectroscopic follow-up 
(SDSS, Szkody et al. 2002; 2dF, Marsh et al. 2002; 
HQS, G\"ansicke et al. 2002)
use different candidate selection criteria but also 
are discovering new CVs 
in the magnitude range of 16 to $\sim$21. The initial estimated space density of CVs 
from these surveys 
do not support the predicted numerous faint CV population.  
However, as we will see below, our survey type and candidate selection are 
based on different methods and we go to fainter magnitude limits. 
Fainter candidate search techniques are in line with theoretical predictions 
for a larger space density of low luminosity CVs (Politano et al. 1998).
In any case, observational identification and study of faint CVs
is needed in order to test our long held ideas about cataclysmic variables and to prove
or eliminate the existence of a large group of faint CVs. The results of such programs 
also have a direct bearing on the white dwarf and binary star production within 
the Galaxy.

We present here our initial search for faint CVs by using candidate sources identified
from two imaging surveys. These two surveys have been undertaken to provide
candidate CVs for follow up spectroscopic confirmation. 
The Faint Sky Variability Survey (Groot et al. 2002) 
used the wide-field camera on the 2.5-m Isaac Newton telescope to provide candidate
sources based on blue color (B-V, V-I) and intrinsic variability on 
time scales of minutes, hours, days
and up to 1 year (see Huber et al. 2002). The FSVS has observed $\sim$20 square degrees 
and presents multi-color data on
point and extended sources from V=16.5 down to V$\sim$24th magnitude. A second survey 
(Davenhall, Clowes, \& Howell 2001;  Clowes et al. 2002) used UK
Schmidt plates in B, R, and H$\alpha$ to identify faint sources which are both blue
(B-R) and likely to be H$\alpha$ sources. These candidates are
of B magnitude 18 to $>$22. Details of the selection criteria for faint CV candidates
from both of these surveys are discussed in the references given above.

\section{Observations}

Our observations were performed using the ESO New Technology Telescope (NTT) and the 
ESO Multi-Mode Instrument (EMMI) on the nights of 9-10 August 2002. 
%During the
%observations, EMMI first provides a
%direct CCD image of the field of view in order to allow the user to unambiguously
%identify the object to place in the slit. These images are saved to disk as part of the
%observing protocol and can be used to provide a brightness estimate for the object 
%at the time of observation. Since our objects were faint (V=18-22), we opted to 
%obtain these CCD images without a filter, yielding a ``white light" observation 
%approximating a broad V+R filter.
%Once the candidate source is marked, dispersing optics are moved into place
%and an integration is started. 
We used EMMI
in RILD (Red Imaging, Low Dispersion) mode with a 360 l/mm grism (grism
\#3) providing
2.3\AA~spectral resolution and wavelength coverage from 3400 to 9200\AA.
The grism efficiency drops blue-ward of 3800\AA~and red-ward of 9000\AA.
Minor second order overlap from 3400 to about 4200\AA~occurred but since the sources 
of interest were
faint and blue, and the grism efficiency in second order at the overlap (red) 
wavelengths is
only 20\% of first order, we used no order sorting filter for these observations.
This decision provided maximum throughput for the spectrograph.

The seeing was excellent on both nights, being between 0.4 to 0.6 arcsec for 
almost all times
and sources. Sources at higher airmass (the northern sky sources had X=1.4-2.0)
had slightly degraded seeing, near 0.8 arcsec. We used a 1.0 arcsec slit for 
all observations, thus 
all the source light entered the spectrograph. We observed
spectrophotometric standards at the start and near the end of each night and bias,
flat, and wavelength calibration arcs were obtained prior to and after each night of
observing. EMMI is a Nasmyth focus instrument, so spectrograph flexure throughout the
night does not occur. Our calibrated spectra have a wavelength scale uncertainty of
0.2 \AA~RMS 
and typical flux uncertainties are of order 10\% over most of the range, 
degrading rapidly near the ends.

Candidate stars were chosen from the FSVS
fields centered near RA 16-17 and DEC +21-28  and RA 21-0 and DEC +27 
(see Groot et al. 2002).
H$\alpha$ candidates were selected from a Schmidt field centered near RA 2:20, DEC -30
(see Clowes et al. 2002). In all, we observed 13 FSVS candidates which were 
known to be ``CV-like" variables and bluer than B-V=1.0, and
V-I=1.2, the CV color range. (see Huber et al. 2002). We also observed 12 H$\alpha$
candidates (Clowes et al. 2002) which were bluer than B-R=1.0 and in fact 
showed no appreciable H$\alpha$ emission 
(except for the discovered CVs and 1 red-shifted
QSO line). This initial survey work was designed to explore the sensitivity of our selection
criteria and has revealed that some emission is probably variable, our R, B, and H$\alpha$
source matching is not yet optimal and/or there is a 
broader main sequence component than we expected. 
From our initial small sample of 25 candidates, we discovered two new CVs, 
two stars with
emission cores in H$\alpha$, and six extragalactic objects. 
The five QSOs observed will
be discussed in detail in Clowes et al. (2002).
Table 1 presents an observing log for the candidates and their identification.

%Table 1
 \begin{table*}[H]
\begin{center}
\caption{Faint CV Candidates}
\begin{tabular}{ccccccc}
 
Source & RA(2000) & DEC(2000) & UT(hrs) & B$^a$ & Int Time (sec) & ID \\
\hline
  & & &   2002 Aug 09 UT & & & \\

FSVSJ1626+2657 & 16:26:27.10 & +26:57:58.8    & 23:34 & 16.4  & 2x180 & sd F,G star \\
FSVSJ1632+2107 & 16:32:36.32 & +21:07:22.4    & 23:55 & 18.6  & 350 &  F,G star \\

  & & &   2002 Aug 10 UT & & & \\

H$\alpha$0231-3140 & 2:31:09.1 & -31:40:34  & 07:53 & 20.5 & 1000 & A,F star \\
H$\alpha$0233-3101 & 2:33;36.0 & -31:01:00  & 06:57 & 19.9 & 800  & A,F star \\
H$\alpha$0236-3058 & 2:36:40.9 & -30:58:22  & 07:42 & 20.4 & 20$^b$ & compact galaxy \\
H$\alpha$0237-3024 & 2:37:08.1 & -30:24:07  & 09:04 & 19.3 & 800 & QSO \\
H$\alpha$0240-2836 & 2:40:46.8 & -28:36:02  & 08:17 & 20.6 & 1000 & F,G star \\
H$\alpha$0241-3045 & 2:41:30.1 & -30:45:37  & 07:17 & 20.1 & 800 & QSO \\
H$\alpha$0241-2923 & 2:41:47.3 & -29:23:40  & 06:07 & 19.4 & 800 & QSO \\
H$\alpha$0242-2802 & 2:42:34.8 & -28:02:43  & 09:57 & 19.0 & 2x700 & CV \\
H$\alpha$0247-3159 & 2:47:07.1 & -31:59:49  & 06:36 & 19.7 & 800 & blue cont. \\
H$\alpha$0249-2748 & 2:49:30.5 & -27:48:09  & 08:41 & 19.4 & 800 & QSO \\
H$\alpha$0251-2837 & 2:51:23.9 & -28:37:52  & 09:24 & 19.4 & 800 & QSO \\
H$\alpha$0254-3224 & 2:54:25.8 & -32:24:29  & 05:31 & 19.9 & 600 & A,F star \\

FSVSJ1722+2733 & 17:22:17.09 & +27:33:21.2    & 01:01 & 20.5  & 800 &  A,F star \\
FSVSJ1722+2723 & 17:22:43.96 & +27:23:55.7    & 00:15 & 20.6  & 2x900 &  CV \\
FSVSJ1722+2714 & 17:22:55.72 & +27:14:14.6    & 00:48 & 19.3  & 350 & F,G star \\ 
FSVSJ1725+2729 & 17:25:09.04 & +27:29:03.9    & 01:55 & 20.9  & 2x1000 & A,F star \\
FSVSJ1727+2737 & 17:27:54.09 & +27:37:43.4    & 02:35 & 18.8  & 250 & F,G star \\
FSVSJ1728+2736 & 17:28:04.43 & +27:36:57.6    & 01:23 & 17.1  & 230 & H$\alpha$ em. cores, narrow
ab. lines \\
FSVSJ2159+2743 & 21:59:18.92 & +27:43:29.8   &  03:40 & 20.7  & 600 &  F,G star \\
FSVSJ2159+2737 & 21:59:51.39 & +27:37:06.2   &  03:58 & 18.0  & 350 &  H$\alpha$ em. cores, narrow
ab. lines \\
FSVSJ2200+2729 & 22:00:44.19 & +27:29:10.6    & 04:22 & 21.4  & 2x1500 & double, red+blue stars \\

 & & &   2002 Aug 11 UT & & & \\

FSVSJ2341+2726 & 23:41:21.52 & +27:26:28.9 & 07:59 & 18.4 & 350 & blue cont., no/weak lines \\
FSVSJ2348+2826 & 23:48:16.01 & +28:26:29.3 & 08:12 & 18.4 & 350 & blue cont., no/weak lines \\
\hline
\end{tabular}
\end{center}
\begin{list}{}{}
\item[$^{\mathrm{a}}$] The B magnitude is from the imaging survey observations.
\item[$^{\mathrm{b}}$] Image only, no spectrum obtained.
\end{list}
   \end{table*}

\section{Discussion}

Both imaging surveys provide accurate positions and magnitudes for the candidates. 
We present finding charts for the
new CVs in Figs. 1 \& 3, our FSVS V-band light curve 
in Fig. 2,
and the EMMI spectra in Fig. 4. Table 2 provides some details
of the properties of the new CVs along with magnitude information obtained from
the imaging surveys.

%Table 2

\begin{table*}[H]
\begin{center}
\caption{New CV Candidate Properties}
\begin{tabular}{cccccl} 
 Source & B  & V & R & I & Spectroscopic Properties\\
\hline

H$\alpha$0242-2802 & 19.0 &  --  & 18.6  & --   &  Double peaked emission lines of H, He, 
                                      Strong He I (5876\AA \\
FSVSJ1722+2723 &  20.6 & 20.4-21.0$^a$  & --  & 19.6   & Strong Em lines of H, He,
$\delta$V20-21\\
\hline
\end{tabular}
\end{center}
\begin{list}{}{}
\item[$^{\mathrm{a}}$] FSVSJ1722+2723 varies on few hour to day 
time scales (see Fig. 2.)
\end{list}
\end{table*}

\subsection{H$\alpha$0242-2802}

H$\alpha$0242-2802 has the B-R color of a mid- to late A star, similar to 
the short period
TOADs, WZ Sge and GO Com. In fact, the spectrum obtained for H$\alpha$0242-2802
(Fig. 4) 
has an appearance reminiscent of WZ Sge (also shown in Fig. 4) 
with strong, double peaked emission lines of H
and He (the central absorption extending to near or below the local continuum for
the He lines and later Balmer lines), and evidence for white dwarf absorption in the 
later Balmer lines. We also note that in H$\alpha$0242-2802 and WZ Sge the Paschen
series lines are in emission as well.
The continuum appearance, the white dwarf absorption, and the emission
line structures in general are indicative of a high inclination, short orbital period, 
low mass transfer CV. A determination of the orbital period for
H$\alpha$0242-2802 is needed and, if similar to WZ Sge as well,
would provide confirmation of these inferences.
Making a tentative guess, based on the assumed similarity with WZ Sge, we take
the M$_V$ of H$\alpha$0242-2802 to be near 12
yielding a z height of 250 pc.

% figure as large as the width of the column
%-------------------------------------------------------------
   \begin{figure}
%   \centering
%   \includegraphics[width=\textwidth,height=5cm]{fsvscv.ps}
\rotatebox{0}{\resizebox{15pc}{!}{\includegraphics[width=8cm]{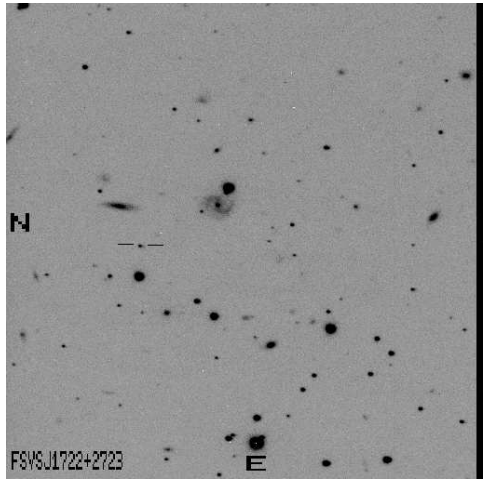}}}
      \caption{Finding chart for FSVSJ1722+2723 made from the EMMI image
obtained prior to the spectroscopic observation. This image is approximately
V+R, is a 20 sec exposure, and is 2.3 arcmin on a side.}
%         \label{FigVibStab}
   \end{figure}
%
%_____________________________________________________________
%-------------------------------------------------------------
   \begin{figure}
%   \centering
%   \includegraphics[width=8cm,height=5cm]{hacv.ps}
\rotatebox{-90}{\resizebox{15pc}{!}{\includegraphics[width=8cm]
{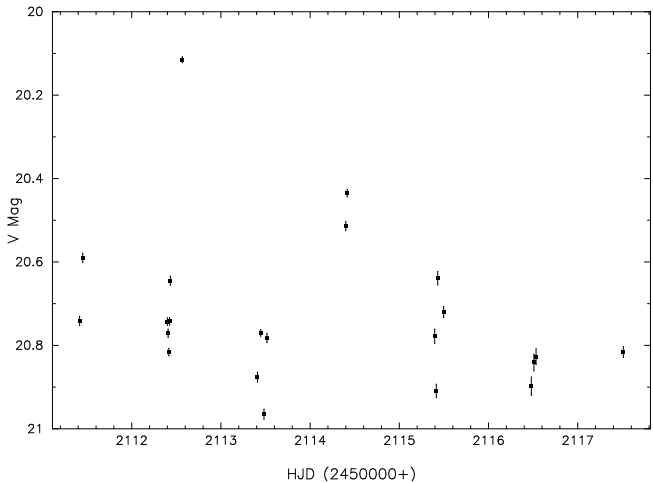}}}
      \caption{FSVS V-band light curve for FSVSJ1722+2723. 
Note the rapid variability
suggestive of a short orbital period.}
%         \label{FigVibStab}
   \end{figure}
%
%_____________________________________________________________

%-------------------------------------------------------------
   \begin{figure}
%   \centering
%   \includegraphics[width=\textwidth,height=5cm]{hacv.ps}
\rotatebox{0}{\resizebox{15pc}{!}{\includegraphics[width=8cm]{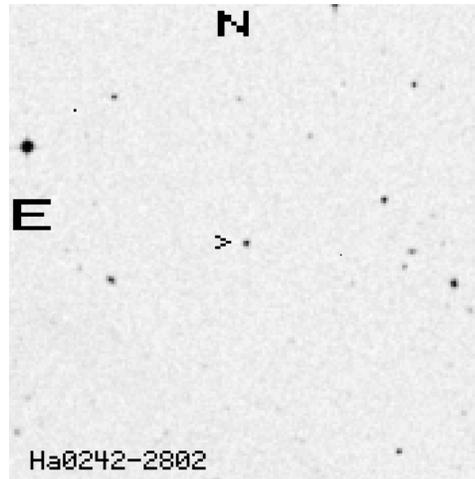}}}
      \caption{Finding chart for H$\alpha$0242-2802 made from the first generation
POSS red plates. The chart is 5 arcmin on a side.}
%         \label{FigVibStab}
   \end{figure}
%
%_____________________________________________________________

%figure
  \begin{figure*}
   \centering
\rotatebox{-90}{\includegraphics[width=12cm]{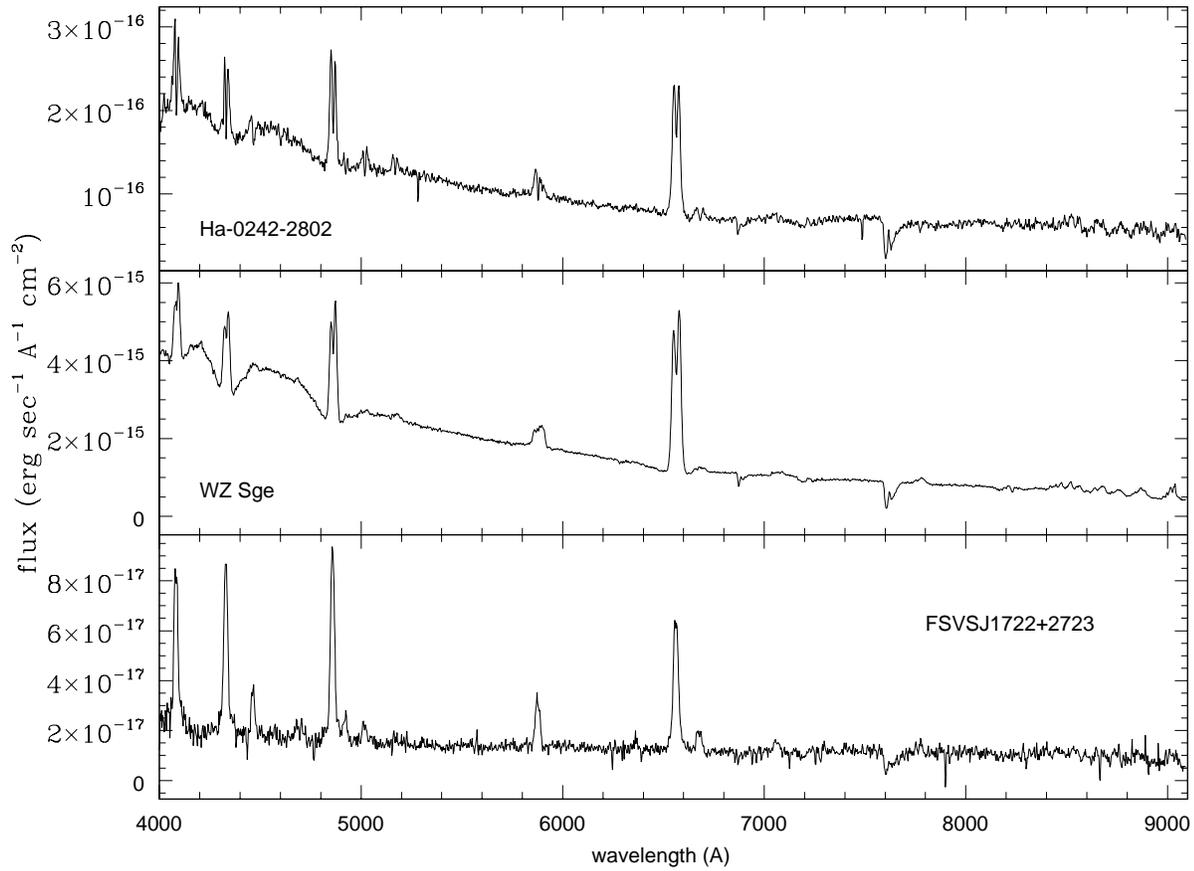}}
   \caption{Discovery spectra of H$\alpha$0242-2802 and FSVSJ1722+2723 obtained 
at the NTT.
            Also shown is the spectrum of the well known TOAD WZ Sge obtained
            with the same instrumental setup on the same night 
            and presented for comparison. Note that H$\alpha$0242-2802 appears very 
            similar to WZ Sge and that FSVSJ1722+2723 has strong He I emission 
            and an inverse H$\alpha$/H$\beta$ decrement.}
    \end{figure*}

\subsection{FSVSJ1722+2723}

FSVSJ1722+2723 is a blue object with B-V=-0.147 and V-I=1.092, similar to the colors of
the nova-like CV, UU Aqr. The emission lines are single peaked (see Fig. 2), He I
(4471, 5876\AA) is fairly strong and the Balmer decrement (H$\alpha$/H$\beta$) 
appears to be inverted. 
The continuum emission rising toward the blue and the high He I
excitation level generally 
indicate a fairly high mass accretion rate. 
However, the weakness of the He II emission and the Balmer reversal are
unusual. If 
FSVSJ1722+2723 is confirmed as a high mass transfer rate
dwarf novae, then we can assume M$_V$$\sim$7-9 which would imply
a large distance to FSVSJ1722+2723, in excess of z=1000 pc. 

\section{Conclusion}

Using our initial small sample of CV discoveries from two different surveys, 
we can estimate
the total number of faint CVs which may be lurking in the remaining bulk of the datasets. We did not explicitly choose the most obvious or brightest 
candidates (given our ability to observe them at the NTT with EMMI) 
from our source lists as a starting point.
Thus, our sample should be relatively unbiased in selection from the candidate
lists but clearly biased
in overall selection criteria as is true of every survey.
The FSVS candidates were chosen as blue, variable sources and we found 1 CV out of 13
sources, $\sim$7\% success. The H$\alpha$ survey candidates, picked by blue color and 
expected H$\alpha$ emission, netted 1 new CV out of 12 candidates, again $\sim$7\%
success. The FSVS has produced a list of $\sim$1200 CV candidates from the entire 
dataset (23 sq. degrees) covering the magnitude range of V=16.5-23 (see Huber et al. 2002)
while the H$\alpha$ survey candidate list
of CVs contains $\sim$500 candidates in three regions 
(RA,DEC's 2:30,-30 and 10:40,+5, +10).
Taking the $\sim$7\% value as indicative of the success rate 
for each survey, we estimate that $\sim$100 additional CVs await discovery within 
our two survey candidate lists. This large number of low luminosity CVs within
our survey areas, 
{\it if in fact realized}, would provide a space density
consistent with the large (undiscovered) population predicted 
by theory (e.g., Howell et al. 2001).
However, many additional candidates from these two imaging surveys 
as well as detailed phase-resolved
spectroscopy must be obtained before such claims can verified.

\begin{acknowledgements}
We wish to thank Jon Willis, Jorge Miranda 
and Karla Aubel for their expert help with the instrument and
observations at La Silla.
\end{acknowledgements}

\end{document}